\documentclass[a4paper]{article}
\usepackage{hyperref}
\usepackage[T1]{fontenc} 
\usepackage[utf8]{inputenc} 
\usepackage{amsmath}         
\usepackage{amssymb}
\usepackage{amsthm}
\usepackage{graphicx} 
\usepackage{bbm}             
\usepackage{bm}              
\usepackage[mathscr]{eucal}  
\usepackage{blindtext}

\title{Violation of Mermin's version of a Bell inequality in a classical statistical model}
\date{\today}
\author{Manfried Faber\footnote{faber@kph.tuwien.ac.at}\\[3mm]
Atominstitut\\
Technische Universität Wien\\
 Operngasse 9, 1040 Vienna, Austria}
\begin{document}
\maketitle
\begin{abstract}
We investigate a classical statistical model and show that Mermin's version of a Bell inequality is violated. We get this violation, if the measurement modifies the ensemble, a feature, which is also characteristic for measurement processes for quantum systems.
\end{abstract}

\section{Introduction}
Important distinctions between classical and Quantum systems are the nonlocal correlations between parts of a quantum system. John Bell formulated this difference in a powerful inequality~\cite{Bell:111654} which rules out local hidden variables as explanations for the measurement outcomes.

Merwin~\cite{Mermin:1981gb} gave a nice example of a Bell-type inequality. It was nicely described by Preskill~\cite{Preskill:1998aa} and Maccone~\cite{Maccone:2013aa}. A non-technical explanation of Bell's inequality was published by Alford~\cite{Alford:2015xpa}. In this article we investigate a classical statistical model and show that Mermin's version of a Bell inequality is violated. We get this violation, if the measurement modifies the ensemble, a feature, which is also characteristic for the measurement process for quantum systems.

Recently, Jaroslaw Duda suggested to investigate the violation of Bell inequalities in a model of maximal entropy random walk~\cite{Duda2009aa}. In the following we follow Duda's idea and present a slightly modified version. We describe an example of a random walk where we follow as close as possible to the presentation of Maccone~\cite{Maccone:2013aa} concerning Mermin's version of a Bell inequality.

\section{Mermin's version of a Bell inequality}
\begin{figure}[h!]
\centering
\includegraphics[scale=0.75]{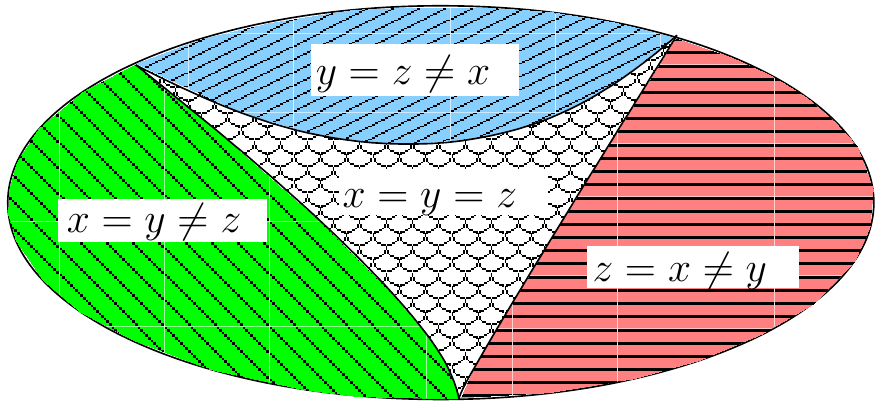}
\caption{The area indicates the probability for the indicated relation between the values $x, y, z$.}
\label{mermin}
\end{figure}

We observe objects $A$ which have three properties. These objects appear always in pairs with equal properties
\begin{equation}\label{pairs}
x_A = x_B,\ y_A = y_B,\ z_A=z_B\quad\textrm{with}\quad x, y, z\in\{0,1\}
\end{equation}
Since we know, that $A$ and $B$ have the same properties we need to check the combination of different properties only. In order not to disturb the measurements we test the first property at $A$ and the second property at $B$. Due to condition~(\ref{pairs}) $(x_A,y_B)$ gives the same result as $(x_B,y_A)$. Therefore, we can omit the indices $A$ and $B$. There are three possibilities for the test of pairs
\begin{equation}\label{3paare}
(x,y), (y,z), (z,x).
\end{equation}
Whatever the probabilities of triples $x, y, z$ are, we get according to Mermin
\begin{equation}\label{Ungleichung}
P_{x=y}+P_{y=z}+P_{z=x}\ge 1,
\end{equation}
as we can easily understand from the diagram in Fig.~\ref{mermin}, where the size of the area is proportional to the probability of the indicated relations.

\begin{figure}[h!]
\centering
\includegraphics{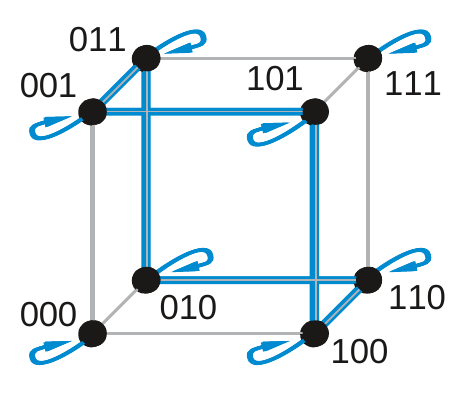}
\caption{Three properties $x$, $y$ and $z$ of objects $A$ and $B$ can take the values $0$ and $1$ only. We  allow for 20 transitions between adjacent sites of the eight $xyz$-triples which are indicated by thick lines. Transitions between adjacent sites can be way and back. Also none of the properties may flip in a time step.}
\label{cube}
\end{figure}

Assuming that the objects can make certain flips of their properties $x, y, z$ we get random walk models. With the aim to finally violate the inequality~(\ref{Ungleichung}) we allow only certain flips of the properties in one time step. In Fig.~\ref{cube} these flips are indicated by blue edges. Only one of the properties can be flipped in one time step, from $0$ to $1$ or from $1$ to $0$. It is also allowed than none of the properties flips. Therefore, there are 20 flips possible.

Using the order of sites
\begin{equation}\label{Ordnung}
(111),(110),(100),(101),(001),(011),(010),(000)
\end{equation}
we can specify the possible flips by the adjacency matrix
\begin{equation}\label{MatM}
M=\begin{pmatrix}
1&0&0&0&0&0&0&0\\
0&1&1&0&0&0&1&0\\
0&1&1&1&0&0&0&0\\
0&0&1&1&1&0&0&0\\
0&0&0&1&1&1&0&0\\
0&0&0&0&1&1&1&0\\
0&1&0&0&0&1&1&0\\
0&0&0&0&0&0&0&1
\end{pmatrix}.
\end{equation}
For the random walk we assumes that all paths are equally probable. The number of paths passing at time $t$ at the eight points we store in a vector with eight numbers
\begin{equation}\label{Folge}
n_t=M^tn_0.
\end{equation}
Since the new distribution depends on the previous only, the process~(\ref{Folge}) is a Markov process.

Starting with $n_0(i)=\delta_{4i}$ we get the sequence
\begin{equation}\begin{aligned}\label{ersteWerte}
&n_0=(0, 0, 0, 1, 0, 0, 0, 0)\\
&n_1=(0, 0, 1, 1, 1, 0, 0, 0)\\
&n_2=(0, 1, 2, 3, 2, 1, 0, 0)\\
&n_3=(0, 3, 6, 7, 6, 3, 2, 0)\\
&n_4=(0,11,16,19,16,11, 8, 0)\\
&n_5=(0,35,46,51,46,35,30, 0)
\end{aligned}\end{equation}
which is soon approaching a constant distribution in the region $R=[2,7]$. If the process started at $t=-\infty$ we get at any finite time the same constant distribution, with the probability $p(i)=1/6$ that a path is arriving in a point $i\in R$. Since we have at every point in the region $R$ either $x=y$ or $y=z$ or $z=x$ and never $x=y=z$ we get for the probabilities $P$ of these paths
\begin{equation}\label{Wxyz}
P_{x=y}+P_{y=z}+P_{z=x}=1,
\end{equation}
in agreement with Mermin's version of a Bell inequality~(\ref{Ungleichung}).

To get a violation of this inequality, we define a measurement process which modifies the ensemble. This measurement process needs one time step. We request, that during this step the measured coordinates can not be changed. Further we never measure all three properties at once since the mearsurement of the third property would change the outcome of the two other properties.

With these assumptions we fulfil Maccones suppositions in~\cite{Maccone:2013aa}, ``that the values of these properties are predetermined (counterfactual definiteness) and not generated by their measurement, and that the determination of the property of one object will not influence any property of the other object (locality).''

\begin{figure}[h!]
\centering
\includegraphics[scale=0.8]{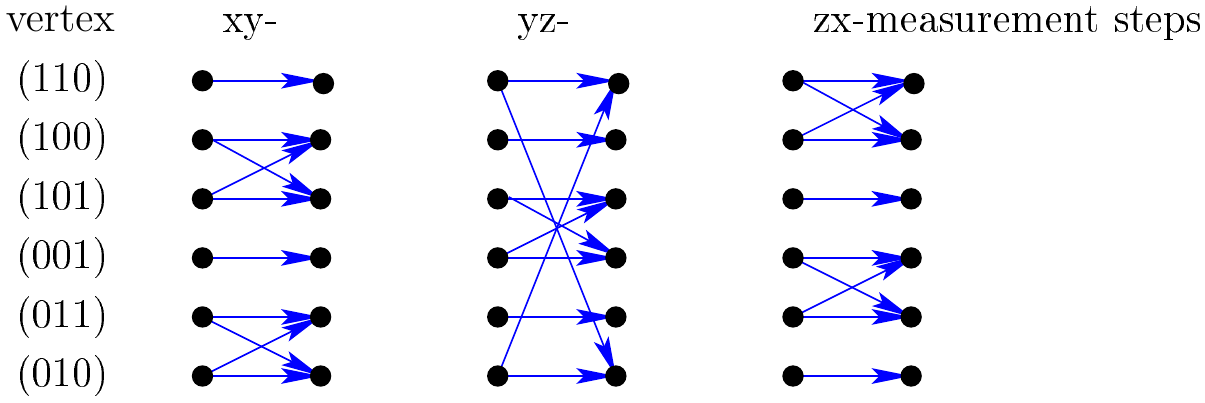}
\caption{An infinite long Markov process according to Eq.~(\ref{Folge}) leads to a uniform distribution of possible paths. Here the measurement steps are shown. In the measurement steps the two measured values are not allowed to change. The third value is not measured and can vary.}
\label{messung}
\end{figure}

Due to the Markov property we need to investigate only the measurement step since we know that an infinite number of steps before the measurement lead to a constant distribution in $R$. We show in Fig.~\ref{messung} the possible paths for $x$ and $y$ measurements. It is important that during an $xy$-measurement $z$ is not observed and can be modified.
Only two of the 10 trajectories in the $xy$-measurement step belong to the results $x=y$ leading to the probality $P_{x=y}=\frac{2}{10}$. In contradiction to the inequality~(\ref{Ungleichung}) we conclude
\begin{equation}\label{Wxyz}
P_{x=y}+P_{y=z}+P_{z=x}=\frac{6}{10}.
\end{equation}

We observe that there are 18 possibile trajectories during one time step. In an $xy$-measurement only 10 of these trajectories are realised and only two of them contribute to the probability $P_{xy}$. We conclude that the modification of the ensemble of paths in the measurement process is the reason for the violation of the Bell inequality~(\ref{Ungleichung}).

Also in quantum mechanics a measurement modifies the state, if it is not done in an eigenstate of the corresponding operator. E.g. if a spin state is in the superposition $\frac{1}{2}|0\rangle+\frac{\sqrt 3}{2}|1\rangle$ of eigenstates of $s_3$, then by a measurement of $s_3$ the state makes with probability $\frac{1}{4}$ a transition to the state $|0\rangle$ or with probability $\frac{3}{4}$ to $|1\rangle$.

\section{Acknowledgement}
I am grateful to Jarek Duda for his kind hospitality at the Jagiellonian University in Krakow and his careful explanations of his maximal entropy random walk model (MERW) violating Mermin's version of a Bell inequality.

\bibliographystyle{utphys}
\bibliography{literatur}

\end{document}